\documentclass[twocolumn]{autart}

\usepackage{graphicx}
\usepackage{amsmath}
\usepackage{amssymb}
\usepackage{xcolor}

\newtheorem{Thm}{Theorem}
\newtheorem{Asu}{Assumption}
\newtheorem{Pro}{Proposition}

\newtheorem{Lem}{Lemma}
\newtheorem{Col}{Corollary}

\newtheorem{Rem}{Remark}

\newcommand{\g}{{\rm g}}
\newcommand{\diag}{{\rm diag}}
\newcommand{\sgn}{{\rm sgn}}
\usepackage{subcaption}

\begin{document}

\begin{frontmatter}

\title{Robust consensus control of second-order uncertain multiagent systems with velocity and input constraints (extended version)} 


\author[gang2]{Gang Wang}\ead{gwang@usst.edu.cn},
\author[zuo]{Zongyu Zuo}\ead{zzybobby@buaa.edu.cn},
\author[CL]{Chaoli Wang}\ead{clclwang@126.com}
\address[gang2]{Institute of Machine Intelligence, University of Shanghai for Science and Technology, Shanghai 200093, China}
\address[zuo]{School of Automation Science and Electrical Engineering, Beihang University, Beijing 100191, China}
\address[CL]{Department of Control Science and Engineering, University of Shanghai for Science and Technology, Shanghai 200093, China}

\begin{keyword}                           
consensus, fixed-time control, multiagent system, uncertain dynamics. 
\end{keyword}

\begin{abstract}  
In this paper, we investigate the consensus problem of second-order multiagent systems under directed graphs. Simple yet robust consensus algorithms that advance existing achievements in accounting for velocity and input constraints, agent uncertainties, and lack of neighboring velocity measurements are proposed. Furthermore, we show that the proposed method can be applied to the consensus control of uncertain robotic manipulators with velocity and control torque constraints. We rigorously prove that the velocity and control inputs stay within prespecified ranges by tuning design parameters \textit{a priori} and that asymptotic consensus can be achieved through Lyapunov functions and fixed-time stability. Simulations are performed for symmetric and asymmetric constraints to show the efficiency of the theoretical findings.
\end{abstract}
\end{frontmatter}

\section{Introduction}
The distributed control of multiagents is an intriguing problem that many researchers have devoted effort to over the past decade. The appeal of the distributed control of multiagents lies in the wide range of potential applications and the technical challenge posed by ensuring global collective behavior through local interactions. One of the major topics in this research area is distributed consensus, which presents an appropriate controller for agents to realize some agreement on their states \cite{francis2016flocking,mei2016distributed,wang2019distributed}. However, most results in the literature focus on the canonical rendezvous or consensus control issue while ignoring the realistic constraints on agent dynamics.

An attempt was made by \cite{ren2008consensus} to address the consensus problem for double-integrator dynamics using bounded control inputs under an undirected graph. By constructing an auxiliary system for each agent, velocity-free consensus strategies were presented for double-integrator dynamics with input constraints in \cite{abdessameud2010consensus}. An event-triggered distributed consensus protocol was designed for single-integrator systems in \cite{yi2019distributed}. State and output synchronization results for both continuous and discrete-time systems subject to input saturation were established by \cite{saberi2022synchronization,liu2018passivity,yang2014global,yang2016periodic}. In addition, a low-gain based synchronization method for multiagent systems with actuator saturation and unknown nonuniform input delay was proposed in \cite{zhang2020semi}. Nevertheless, the velocity constraint is not considered in the above double-integrator work \cite{ren2008consensus,abdessameud2010consensus}. Recently, the distributed consensus problem for multiagent systems with velocity and control input constraints was considered in \cite{lin2017distributed}. More recently, in contrast to the discrete-time system \cite{lin2017distributed}, consensus in a continuous-time system with velocity and control input constraints was attempted by \cite{fu2019consensus}. However, it is worth noting that this pioneering work does not take into account any system uncertainties other than the need for additional neighboring velocity information. A real-life example of a consensus control scenario with input and velocity constraints, as well as model uncertainty, is the cooperative adaptive cruise control of a platoon of vehicles aimed at maintaining a stable formation. The velocity constraints of the vehicles are established to enhance both traffic flow uniformity and operational safety, while the engine dynamics is uncertain and subject to saturation. These observations provide the main motivation of this paper.

The current work takes a step towards addressing the consensus problem of second-order multiagent systems subject to velocity and control input constraints and dynamics uncertainties. To the best of our knowledge, this work is the first to deal with both issues simultaneously. In the absence of neighboring velocities, a new simple, yet robust control framework is tailored under a general directed graph. Furthermore, we apply the proposed framework to address the consensus control of uncertain robotic manipulators with velocity and torque constraints. Theoretical and simulation verifications of the developed controller assuring that asymptotic consensus can be achieved while the velocity and input constraints are not transgressed at all times are carefully studied.

\textit{Notation:} For any $\alpha>0$, we define $\lfloor x\rceil^{\alpha}=|x|^{\alpha}\sgn(x)$, where $\sgn(\cdot)$ denotes the standard signum function. $0_n$ and $1_n$ represent the $n-$vector of all zeros and all ones, respectively, and $\max\{x_i\}$ is the maximum value among $x_1,\dots,x_n$. A square matrix is a nonsingular $\mathcal{M}$-matrix if all its eigenvalues have positive real parts and all its off-diagonal entries are nonpositive.

\section{Problem formulation and preliminaries}
\subsection{Graph theory}
The interaction topology between agents is represented by a weighted directed graph $\mathcal{G}=(\mathcal{V},\mathcal{E})$, where $\mathcal{V}=\{1,\dots,n\}$ denotes the node set and $\mathcal{E}\subseteq \mathcal{V}\times \mathcal{V}$ is the edge set. An edge $(i,j)\in\mathcal{E}$ means that the information of node $i$ is available to node $j$. A directed path from node $i_1$ to node $i_p$ is a sequence of ordered edges in the form $(i_m,i_{m+1}),$ $m=1,\dots,p-1$. We say that a directed graph is strongly connected if there exists a path in $\mathcal{G}$ between any two nodes. A directed tree is a directed graph in which every node has exactly one parent except for one node, which has directed paths to every other node. The adjacency matrix $\mathcal{A}=[a_{ij}]\in R^{n\times n}$ is defined by $a_{ij}>0$ if $(j,i)\in \mathcal{E}$ and $a_{ij}=0$ otherwise. Define $d_i=\sum_{j=1}^na_{ij}$ as the in-degree of node $i$ and $\mathcal{D}=\diag\{d_1,\dots,d_n\}$. The Laplacian matrix is defined as $\mathcal{L}=\mathcal{D}-\mathcal{A}$.
\begin{Lem} \cite{mei2016distributed} \label{Lem1}
If $\mathcal{G}$ is strongly connected, there exists a vector $\omega=[\omega_1,\dots,\omega_n]$ with $\omega_i>0$ for all $i=1,\dots,n$ such that $\omega\mathcal{L}=0_n^T$. Define the matrix $\hat{\mathcal{L}}=W\mathcal{L}+\mathcal{L}^TW$, where $W=\diag\{\omega_1,\dots,\omega_n\}$. Then, $\hat{\mathcal{L}}$ is positive semidefinite, and the null space of $\hat{\mathcal{L}}$ is $\{\rho1_n: \rho\in R\}$.
\end{Lem}
\subsection{Problem formulation}
Consider a second-order uncertain multiagent system described by the following form:
\begin{equation}\label{agent}
\begin{array}{cll}
\dot{x}_{i}&=&v_i,\\ \dot{v}_{i}&=&b_i(t)u_i+\tau_i(t),
\end{array}
\end{equation}
where $i=1,\dots,n$, $x_i\in R$, $v_i\in R$, and $u_i\in R$ are the position, velocity, and control input of the $i$th agent, respectively, and $b_i(t)>0$ and $\tau_i(t)$ may be taken as unmodelled dynamics, including model uncertainties and unknown disturbances. 
Considering safety specifications and actuator saturation, both the velocity and the control input need to meet the following constraints:
\begin{equation}\label{constraints_vel}
|v_i(t)|\leq v_{\max}, |u_i(t)|\leq u_{\max}, \forall t\geq 0,
\end{equation}
where $v_{\max}$ and $u_{\max}$ are given positive constants and are available to each agent. Throughout this paper, the initial velocity $v_i(0)$ is always assumed to satisfy the given constraint.

The control goal is to design a distributed control law without requiring the velocity of neighboring agents, which enables all agents to reach consensus on the position state, i.e., $x_i(t)-x_j(t) \rightarrow 0$ as $t\rightarrow\infty$ for all $i,j=1,\dots,n$, and guarantees that the velocity and input constraints given by (\ref{constraints_vel}) are satisfied at all times.

\begin{Asu}\label{Asu2}
There exist known positive constants $b_{\min}$ and $\tau_{\max}$ such that $b_i(t)\geq b_{\min}$ and $|\tau_i(t)|\leq \tau_{\max}$ for all $i=1,\dots,n$. In addition, $b_{\min}$ and $\tau_{\max}$ need to satisfy $\tau_{\max}<b_{\min}u_{\max}$ to bypass the loss of controllability issue. 
\end{Asu}
\begin{Asu}\label{Asu1}
The directed graph $\mathcal{G}$ among the multiagent systems is strongly connected.
\end{Asu}

\section{Controller design and main results}\label{sec3}
\subsection{Symmetric velocity constraints}
For notational convenience, we define $\eta_i=\sum_{j=1}^na_{ij}(x_i-x_j)$ and $\zeta_i=\sum_{j=1}^na_{ij}(v_{i}-v_{j})$. We propose the following distributed robust control law:
\begin{equation}\label{u}
u_i =-\frac{(u_{\max}-\alpha_i)\sigma(e_i)+\lambda_i\alpha_i\sgn(e_i)}{\lambda_i},
\end{equation}
where $e_i=v_i-r_i$ denotes the velocity tracking error, $r_i$ is the reference velocity of agent $i$ designed as
\begin{equation}\label{r}
r_i=-m\tanh\big(\frac{k_i\eta_i}{m}\big), 
\end{equation}
$\lambda_i=z_i^{\gamma_i}$ with $\gamma_i>1$, and $\sigma(e_i)$ is a nonlinear saturated function with
\begin{equation}\label{sigma}
\sigma(e_i)=\left\{
\begin{array}{ll}
\lambda_i, &{\rm ~if~}  e_i\geq z_i\\
\lfloor e_i\rceil^{\gamma_i}, &{\rm ~if~} -z_i<e_i<z_i\\
-\lambda_i, &{\rm ~if~} e_i\leq -z_i.\\
\end{array}
\right.
\end{equation}
Here, $m$, $\alpha_i$, $z_i$, and $k_i$ are positive constants satisfying
\begin{equation}\label{parameter}
\begin{array}{c}
m<v_{\max}, (\tau_{\max}/b_{\min})<\alpha_i<u_{\max},\\
 z_i\leq v_{\max}-m, k_i<(b_{\min}\alpha_i-\tau_{\max})/(2d_iv_{\max}).
 \end{array}
\end{equation}
Now, the first major result of this paper is stated and proven.
\begin{Pro}\label{Pro1}
Given any positive $v_{\max}$ and $u_{\max}$, the constraint (\ref{constraints_vel}) is not breached for our proposed controller (\ref{u}) with (\ref{r}) and (\ref{parameter}).
\end{Pro}
\textit{Proof.} It follows from (\ref{u}) that $u_i=-((u_{\max}-\alpha_i)\sigma(e_i))/\lambda_i+\alpha_i$ for $e_i<0$. Noting $-\lambda_i\leq\sigma(e_i)\leq0$, we obtain $\alpha_i \leq u_i\leq u_{\max}.$ For $e_i>0$, it follows that $0\leq\sigma(e_i)\leq\lambda_i$ and $u_i=-((u_{\max}-\alpha_i)\sigma(e_i))/\lambda_i-\alpha_i$, leading to $-u_{\max} \leq u_i\leq -\alpha_i.$ When $e_i=0$, $u_i=0$. Consequently, it holds that $|u_i(t)|\leq u_{\max}$ for all $t\geq 0$.

We now demonstrate that $|v_i(t)|\leq v_{\max}$. Noting (\ref{r}), we obtain that $|r_i(t)|\leq m$. When $v_i=v_{\max}$, it follows from (\ref{parameter}) that $e_i=v_i-r_i\geq v_{\max}-m\geq z_i>0,$ which together with (\ref{u}) and (\ref{sigma}) implies $\sigma(e_i)=\lambda_i$ and $u_i=-u_{\max}$. When $v_i=-v_{\max}$, we have $e_i=v_i-r_i\leq -v_{\max}+m\leq -z_i<0,$ resulting in $u_i=u_{\max}$. 
Therefore, we can obtain
\begin{equation*}
\dot{v}_i=\left\{
\begin{array}{ll}
b_iu_i+\tau_i,& {\rm if~~} |v_i|<v_{\max}\\
-b_iu_{\max}+\tau_i,& {\rm if~~} v_i=v_{\max}\\
b_iu_{\max}+\tau_i,& {\rm if~~} v_i=-v_{\max}.
\end{array}
\right.
\end{equation*}
Define the Lyapunov function candidate $P_i=v_i^2$. Its time derivative satisfies $\dot{P}_i=2v_i\dot{v}_i$. The following three cases are considered:
    \begin{enumerate}
        \item If $|v_i|<v_{\max}$, the statement $|v_i|\leq v_{\max}$ holds.
        \item If $v_i=v_{\max}$, we obtain $\dot{P}_i=2v_{\max}(-b_iu_{\max}+\tau_i)$. By Assumption \ref{Asu2}, it holds that $\dot{P}_i(t)\leq-2v_{\max}(b_{\min}u_{\max}-\tau_{\max})\leq 0$. Thus, $P_i$ is nonincreasing, and accordingly, we obtain $|v_i|\leq v_{\max}$.
        \item If $v_i=-v_{\max}$, we have $\dot{P}_i=-2v_{\max}(b_iu_{\max}+\tau_i)\leq-2v_{\max}(b_{\min}u_{\max}-\tau_{\max})\leq 0$, which also implies $|v_i|\leq v_{\max}$.
    \end{enumerate}
Therefore, since $v_i$ is continuous and $|v_i(0)|\leq v_{\max}$, we conclude that $|v_i(t)|\leq v_{\max}$ for all $t\geq 0$.

\begin{Thm}\label{Thm1}
Consider a multiagent system of $n$ agents (\ref{agent}) satisfying Assumptions \ref{Asu2}-\ref{Asu1}. The proposed distributed robust control law (\ref{u}) with (\ref{r}) and (\ref{parameter}) ensures that the agents achieve asymptotic consensus and the agent velocity converges to zero, i.e., $\lim_{t\rightarrow\infty}(x_i(t)-x_j(t))=0$ and $\lim_{t\rightarrow\infty}v_i(t)=0$ for all $i,j=1,\dots, n$.
\end{Thm}
\textit{Proof.} We begin the proof by considering the time derivative of $r_i$ in (\ref{r}), which satisfies $\dot{r}_i=-k_i(1-\tanh^2((k_i\eta_i)/m))\zeta_i.$ It has been proven in Proposition \ref{Pro1} that $|v_i(t)|\leq v_{\max}$. This indicates that $|\zeta_i(t)|\leq 2d_iv_{\max}$ for all $t\geq 0$. Observing from (\ref{parameter}) that $k_i<(b_{\min}\alpha_i-\tau_{\max})/(2d_iv_{\max})$, we further have $|\dot{r}_i(t)|< b_{\min}\alpha_i-\tau_{\max}$. The time derivative of $e_i$ is obtained as 
$\dot{e}_i=b_iu_i+\tau_i-\dot{r}_i$. For $e_i\geq z_i$, we obtain from (\ref{u}) that $u_i=-u_{\max}$. Therefore, we have
\begin{equation*}
\dot{e}_i=-b_iu_{\max}+\tau_i-\dot{r}_i<-b_{\min}(u_{\max}-\alpha_i)<0,
\end{equation*}
where $\alpha_i<u_{\max}$ is applied to obtain the last inequality. For $e_i\leq -z_i$, we obtain that $u_i=u_{\max}$, which implies
\begin{equation*}
\dot{e}_i=b_{\min}u_{\max}+\tau_i-\dot{r}_i>b_{\min}(u_{\max}-\alpha_i)>0.
\end{equation*}
Note from $|v_i(0)|\leq v_{\max}$ that $|e_i(0)|\leq |v_i(0)|+|r_i(0)|\leq v_{\max}+m$. Thus, we have $|e_i(t)|\leq z_i$ for all $t\geq T_{i,1}$, where $T_{i,1}\leq (v_{\max}+m-z_i)/(b_{\min}(u_{\max}-\alpha_i))$. For $t\geq T_{i,1}$, it holds that $\sigma(e_i)=\lfloor e_i\rceil^{\gamma_i}$, yielding the conclusion that the dynamics of $e_i$ takes the following form 
\begin{equation}\label{dot_e}
\dot{e}_i=-\frac{b_i(u_{\max}-\alpha_i)\lfloor e_i\rceil^{\gamma_i}}{\lambda_i}-b_i\alpha_i\sgn(e_i)+\tau_i-\dot{r}_i.
\end{equation}
Noting $|\dot{r}_i(t)|\leq 2k_id_iv_{\max}$, we consider the candidate Lyapunov function $H_i$ as $H_i=\frac{1}{2}e_i^2$, whose time derivative along (\ref{dot_e}) satisfies
\begin{eqnarray}
\dot{H}_i&=&e_i(-\frac{b_i(u_{\max}-\alpha_i)\lfloor e_i\rceil^{\gamma_i}}{\lambda_i}-b_i\alpha_i\sgn(e_i)+\tau_i-\dot{r}_i)\nonumber\\
 &\leq & -(\frac{b_{\min}(u_{\max}-\alpha_i)}{\lambda_i} |e_i|^{\gamma_i+1}+\mu_i|e_i|)\nonumber\\
&\leq & -(\frac{b_{\min}(u_{\max}-\alpha_i)}{\lambda_i} (2H_i)^{\frac{\gamma_i+1}{2}}+\mu_i(2H_i)^{\frac{1}{2}}),\label{dot_h}
\end{eqnarray}
where $\mu_i=b_{\min}\alpha_i-\tau_{\max}-2k_id_iv_{\max}$. The application of Lemma 2 in \cite{zuo2020distributed} to (\ref{dot_h}) ensures the fixed-time convergence of $e_i$, i.e., $e_i=0$ for all $t\geq T_{i,1}+T_{i,2}$, where the settling time function $T_{i,2}$ is bounded by $$T_{i,2}\leq \frac{\lambda_i}{2^{\frac{\gamma_i-1}{2}}b_{\min}(u_{\max}-\alpha_i)(\gamma_i-1)}+\frac{1}{2^{-\frac{1}{2}}\mu_i}.$$ Consequently, for $t\geq T$ with $T=\max\{T_{i,1}+T_{i,2}\}$, the evolution of $x_i$ for all $i=1,\dots,n$ can be written as $\dot{x}_i=r_i=-m\tanh((k_i\eta_i)/m).$ 

It rests to prove that asymptotic consensus can be reached. Motivated by \cite{ren2008consensus,fu2019consensus}, we consider the following Lyapunov function candidate $$V=\sum_{i=1}^n (\omega_i/k_i)\ln(\cosh((k_i\eta_i)/m)),$$
in which $\omega_i$ is defined in Lemma \ref{Lem1}. By recalling $\dot{x}_i=-m\tanh(\xi_i)$ for $t\geq T$, we obtain that for all $t\geq T$ 
\begin{equation*}
\begin{array}{lll}
\dot{V}&=&\sum_{i=1}^n\frac{\omega_i\dot{\eta}_i}{m}\tanh(\xi_i)\\
&=&\sum_{i=1}^n\frac{\omega_i}{m}(\sum_{j=1}^na_{ij}(\dot{x}_i-\dot{x}_j))\tanh(\xi_i)\\
&=&-\sum_{i=1}^n\omega_i(\sum_{j=1}^na_{ij}(\tanh(\xi_i)-\tanh(\xi_j))\tanh(\xi_i)\\
&=&-\tanh^T(\xi)W\mathcal{L}\tanh(\xi)=-\frac{1}{2}\tanh^T(\xi)\hat{\mathcal{L}}\tanh(\xi),
\end{array}
\end{equation*}
where $\xi_i=(k_i\eta_i)/m$, $\hat{\mathcal{L}}$ is defined in Lemma \ref{Lem1}, and $\tanh(\xi)=[\tanh(\xi_1),\dots,\tanh(\xi_n)]^T$. By Assumption \ref{Asu1} and Lasalle’s invariance principle, we can conclude that $\xi$ goes to the largest invariant set $\{\xi:\xi=\rho1_n, \rho\in R\}$ as $t\rightarrow\infty$. Noting from Lemma \ref{Lem1} that 
\begin{equation*}
\bar{\omega}\xi=\bar{\omega}\diag\big\{\frac{k_1}{m},\dots,\frac{k_n}{m}\big\}\mathcal{L}[x_1,\dots,x_n]^T=0,
\end{equation*}
where $\bar{\omega}=[(\omega_1/k_1),\dots,(\omega_n/k_n)]$, we have $\rho\bar{\omega} 1_n=0$. This coupled with $\omega_i>0$ and $k_i>0$ gives $\rho=0.$  As a result, we obtain $\lim_{t\rightarrow\infty}\xi_i(t)=0$, $\lim_{t\rightarrow\infty}\eta_i(t)=0$, and $\lim_{t\rightarrow\infty}(x_i(t)-x_j(t))=0$ for all $i,j=1,\dots,n$. Recalling (\ref{r}), we further have $\lim_{t\rightarrow\infty}v_i(t)=0.$ Here we complete the proof.

\subsection{Asymmetric velocity constraints}
The distributed robust control law (\ref{u}) with (\ref{r}) and (\ref{parameter}) addresses the symmetric velocity constraint. In the following, we show that our proposed robust controller can also be applied to address asymmetric velocity constraints. More precisely, we consider the following constraints:
\begin{equation}\label{constraints_2}
v_{\min}\leq v_i(t)\leq v_{\max}, |u_i(t)|\leq u_{\max}, \forall t\geq 0, 
\end{equation}
where $v_{\min}$ and $v_{\max}$ with $v_{\min}<v_{\max}$ are arbitrary velocity constraints, and $u_{\max}>0$ is the control input constraint. We can design the reference velocity $r_i$ as 
\begin{equation}\label{r_new}
r_i=v_r-m\tanh\big(\frac{k_i\eta_i}{m}\big), 
\end{equation}
where $v_r=(v_{\max}+v_{\min})/2$. Then, the positive constants $m$, $\alpha_i$, $z_i$, and $k_i$ in (\ref{u}) and (\ref{r_new}) require to  satisfy
\begin{eqnarray}\label{parameter_new}
&(\tau_{\max}/b_{\min})<\alpha_i<u_{\max},\nonumber\\  &m<(v_{\max}-v_{\min})/2,
z_i\leq (v_{\max}-v_{\min})/2-m,\nonumber\\
 &k_i<(b_{\min}\alpha_i-\tau_{\max})/(d_i(v_{\max}-v_{\min})).
\end{eqnarray}
\begin{Thm}\label{Thm2}
Consider a multiagent system of $n$ agents (\ref{agent}) satisfying Assumptions \ref{Asu2}-\ref{Asu1}. The proposed distributed robust control law (\ref{u}) with (\ref{r_new}) and (\ref{parameter_new}) ensures that the agents achieve asymptotic consensus and the agent velocity converges to $v_r$, i.e., $\lim_{t\rightarrow\infty}(x_i(t)-x_j(t))=0$ and $\lim_{t\rightarrow\infty}v_i(t)=v_r$ for all $i,j=1,\dots, n$, and that the constraints of agent velocity and control input given by (\ref{constraints_2}) are never violated.
\end{Thm} 
\textit{Proof.} Following a similar proof to Proposition \ref{Pro1}, we can conclude $|u_i(t)|\leq u_{\max}$ for all $t\geq 0$. When $v_i=v_{\max}$, it follows from (\ref{r_new}) and (\ref{parameter_new}) that $$e_i=v_i-r_i\geq v_{\max}-v_r-m\geq z_i>0,$$ which together with (\ref{u}) implies $\sigma(e_i)=\lambda_i$ and $u_i=-u_{\max}$. When $v_i=v_{\min}$, we have $$e_i=v_i-r_i\leq v_{\min}-v_r+m\leq -z_i<0,$$ resulting in $u_i=u_{\max}$. To establish $v_{\min}\leq v_i(t)\leq v_{\max}$, it is sufficient to prove that $P_i(t)=(v_i(t)-v_r)^2\leq (v_{\max}-v_{\min})^2/4$ for all $t\geq 0$. By Assumption \ref{Asu2}, we note that the following inequality holds for all $v_i(t)=v_{\max}$ or $v_i(t)=v_{\min}$, $\dot{P}_i\leq -\frac{1}{2}(v_{\max}-v_{\min})(b_{\min}u_{\max}-\tau_{\max})\leq 0,$
which indicates that $v_{\min}\leq v_i(t)\leq v_{\max}$ for all $t\geq 0$ as long as $v_{\min}\leq v_i(0)\leq v_{\max}$. 

Next, we show that asymptotic consensus can be realized. Since $v_{\min}\leq v_i(t)\leq v_{\max}$, we obtain $|\zeta_i(t)|\leq d_i(v_{\max}-v_{\min})$, which together with (\ref{parameter_new}) implies $|\dot{r}_i(t)|< b_{\min}\alpha_i-\tau_{\max}$. Following steps similar to Theorem \ref{Thm1}, there exists a bounded settling time function $T$ such that for $t\geq T$, the dynamics of $x_i$ for all $i=1,\dots,n$ can be obtained as $\dot{x}_i=v_r-m\tanh(\xi_i)$. Consider the following Lyapunov function candidate $V=\sum_{i=1}^n (\omega_i/k_i)\ln(\cosh(\xi_i)).$ The derivative of $V$ is given by $\dot{V}=-\tanh^T(\xi)\hat{\mathcal{L}}\tanh(\xi)/2,$  
and we can conclude that $\lim_{t\rightarrow\infty}(x_i(t)-x_j(t))=0.$ As an immediate result, we can obtain from (\ref{r_new}) that $\lim_{t\rightarrow\infty}r_i(t)=v_r$ and $\lim_{t\rightarrow\infty}v_i(t)=v_r.$

Note that the control parameter $k_i$ depends on the in-degree $d_i$ of each agent. To avoid this requirement, we propose to incorporate a first-order filter into the controller design. Selecting Theorem 1 as an example, we redesign the reference velocity $r_i$ in (4) of agent $i$ as
\begin{equation}\label{E1}
\begin{array}{lll}
r_i&=&-m\tanh\big(\frac{k_i(x_i-\hat{x}_i)}{m}\big),\\
\dot{\hat{x}}_i&=&-m\tanh(\sum_{j=1}^na_{ij}(\hat{x}_i-x_j)),
\end{array}
\end{equation}
where $\hat{x}_i$ is considered the estimate of the consensus value for the $i$th agent. The control parameters only need to satisfy:
\begin{eqnarray}\label{E2}
&m<v_{\max}, (\tau_{\max}/b_{\min})<\alpha_i<u_{\max},z_i\leq v_{\max}-m,\nonumber\\
&k_i<(b_{\min}\alpha_i-\tau_{\max})/(2v_{\max}).
\end{eqnarray}
Careful observation of (\ref{E2}) reveals that the selection of control parameters is completely independent of the communication graph. Accordingly, we obtain the following result. 

\begin{Col}
Consider a multiagent system of $n$ agents (\ref{agent}) satisfying Assumptions \ref{Asu2}-\ref{Asu1}. The proposed distributed robust control law (\ref{u}) with (\ref{E1}) and (\ref{E2}) ensures that the agents achieve asymptotic rendezvous and the agent velocity converges to zero, i.e., $\lim_{t\rightarrow\infty}(x_i(t)-x_j(t))=0$ and $\lim_{t\rightarrow\infty}v_i(t)=0$ for all $i,j=1,\dots, n$.
\end{Col}

\noindent\textit{Proof.} Following the same steps in Proposition \ref{Pro1} and Theorem \ref{Thm1}, it can be shown that the constraints on agent velocity and control input are not violated under the control law (\ref{u}) with (\ref{E1}) and (\ref{E2}) and that there exists a bounded settling time function $T$ such that for $t\geq T$, $\dot{x}_i=-m\tanh\big(\frac{k_i(x_i-\hat{x}_i)}{m}\big)$. Next, we prove that asymptotic consensus can be realized.

Define the augmented vector $$\bar{x}=[x_1,\dots,x_n,\hat{x}_1,\dots,\hat{x}_n]^T\in R^{2n}$$ and the function vector $$\tanh(y)=[\tanh(y_1),\dots,\tanh(y_{2n})]^T\in R^{2n}$$ for all $y=[y_1,\dots,y_{2n}]^T\in R^{2n}$. Recalling (\ref{E1}), one can verify that $\bar{x}$ satisfies 
\begin{equation}\label{E3}
\dot{\bar{x}}=-m\tanh(\bar{\mathcal{L}}\bar{x})
\end{equation} 
with $K=\diag\{k_1/m,\dots,k_n/m\}$ and 
\begin{equation}\label{E4}
\bar{\mathcal{L}}=
\left[
\begin{array}{cc}
KI_n~ & -KI_n\\
-\mathcal{A}~ & \mathcal{D}
\end{array}
\right],
\end{equation}
where matrices $\mathcal{A}$ and $\mathcal{D}$ are, respectively, the adjacency matrix and diagonal in-degree matrix of $\mathcal{G}$. Matrix $\bar{\mathcal{L}}$ takes the form of a Laplacian matrix since $\bar{\mathcal{L}}1_{2n}=0_{2n}$, all the diagonal entries of $\bar{\mathcal{L}}$ are nonnegative, and all its off-diagonal entries are nonpositive. Therefore, system (\ref{E3}) can be regarded as a system consisting of $2n$ agents that are interconnected according to the augmented graph $\bar{\mathcal{G}}=\{\bar{\mathcal{V}},\bar{\mathcal{E}}\}$, where node $i$ in $\bar{\mathcal{G}}$ denotes the $i$th element of $\bar{x}$. Recalling that the communication graph $\mathcal{G}$ is strongly connected and noting (\ref{E4}), we obtain that there exists a directed path from node $i$ to node $j$ in $\bar{\mathcal{G}}$ for all $i=1,\dots,n$, $j=n+1,\dots,2n$. In addition, since $x_i$ obtains $\hat{x}_i$, there is a directed path from node $i+n$ to node $i$ in $\bar{\mathcal{G}}$ for all $i=1,\dots,n$. Consequently, we conclude that there must exist a directed path for each distinct pair of nodes in $\bar{\mathcal{G}}$, indicating that $\bar{\mathcal{G}}$ is also strongly connected. 

According to Lemma \ref{Lem1}, there exists a vector $\bar{\omega}=[\bar{\omega}_1,\dots,\bar{\omega}_{2n}]$ with $\bar{\omega}_i>0$ for all $i=1,\dots,2n$ such that $\bar{\omega}\bar{\mathcal{L}}=0_{2n}^T$. Consider the following Lyapunov function
\begin{equation*}
V=\sum_{i=1}^{2n}\bar{\omega}_i\ln(\cosh(\xi_i)),
\end{equation*}
where $\xi_i$ is the $i$th element of $\xi$ with $\xi=\bar{\mathcal{L}}\bar{x}$. The time derivative of $V$ is given by 
\begin{equation*}
\dot{V}=-\frac{1}{2}\tanh(\xi)^T\hat{\bar{\mathcal{L}}}\tanh(\xi),
\end{equation*}
where $\hat{\bar{\mathcal{L}}}=\bar{W}\bar{\mathcal{L}}+\bar{\mathcal{L}}^T\bar{W}$ is positive semidefinite with $\bar{W}=\diag\{\bar{\omega}_1,\dots,\bar{\omega}_{2n}\}$. By Lasalle's invariance principle, we obtain $\lim_{t\rightarrow\infty}(\hat{x}_i(t)-x_j(t))=0$ and $\lim_{t\rightarrow\infty}(x_i(t)-\hat{x}_i(t))=0$, and $\lim_{t\rightarrow\infty}(x_i(t)-x_j(t))=0$ for all $i,j=1,\dots,n$.

\subsection{Extension to general directed graphs}
In the above subsection, we investigated the consensus problems with the strongly connected graph. 
The main purpose of this subsection is to extend the obtained results to more general directed graphs having a spanning tree. We introduce the following nonlinear saturation function $\varrho_i: R\rightarrow R$ as:
\begin{equation}\label{varrho}
\varrho_i(s)=\left\{
\begin{array}{ll}
m, &{\rm if~}  s\geq \frac{4m}{3k_i}\\
2k_is-\frac{3}{4m}k_i^2s^2-\frac{m}{3}, &{\rm if~} \frac{2m}{3k_i}\leq s<\frac{4m}{3k_i}\\
k_is, &{\rm if~} -\frac{2m}{3k_i}\leq s<\frac{2m}{3k_i}\\
2k_is+\frac{3}{4m}k_i^2s^2+\frac{m}{3}, &{\rm if~} -\frac{4m}{3k_i}\leq s<-\frac{2m}{3k_i}\\
-m, &{\rm if~}  s\leq -\frac{4m}{3k_i}
\end{array}
\right.
\end{equation}
where $m$ and $k_i$ are positive design parameters. As can be directly checked, $|\varrho_i(s)|\leq m$ and $|\dot{\varrho}_i(s)|\leq k_i$ for all $s\in R$. If the reference velocity $r_i$ in (\ref{r_new}) is redesigned as 
\begin{equation}\label{r_R2_new}
r_i=v_r-\varrho_i(\eta_i),
\end{equation}
we obtain the following result. 
\begin{Thm}\label{Thm_new}
Consider a multiagent system of $n$ agents (\ref{agent}) satisfying Assumption \ref{Asu2} and suppose that $\mathcal{G}$ contains a spanning tree. The proposed distributed robust control law (\ref{u}) with (\ref{parameter_new}) and (\ref{r_R2_new}) ensures that the agents achieve asymptotic consensus and the agent velocity converges to $v_r$, i.e., $\lim_{t\rightarrow\infty}(x_i(t)-x_j(t))=0$ and $\lim_{t\rightarrow\infty}v_i(t)=v_r$ for all $i,j=1,\dots, n$, and that the constraints of agent velocity and control input given by (\ref{constraints_2}) are never violated.
\end{Thm}
\textit{Proof.} By mimicking the arguments used in the proof of Theorems \ref{Thm1}-\ref{Thm2}, it can be shown that the constraints on agent velocity and control input are not violated under the control law (\ref{u}) with (\ref{parameter_new}) and (\ref{r_R2_new}) and that there exists a bounded settling time function $T$ such that for $t\geq T$, $\dot{x}_i=v_r-\varrho_i(\eta_i)$. Next, we prove that asymptotic consensus can be realized under the directed graph $\mathcal{G}$ with a spanning tree.

Inspired by  \cite{mei2016distributed}, we assume that the Laplacian matrix $\mathcal{L}$ has the following Perron-Frobenius standard form:
\begin{equation}\label{Perron}
\mathcal{L}=
\left[
\begin{array}{cccc}
\mathcal{L}_{11}&0&\dots&0\\
\mathcal{L}_{21}&\mathcal{L}_{22}&\dots&0\\
\vdots&\vdots&\ddots&\vdots\\
\mathcal{L}_{\kappa1}& \mathcal{L}_{\kappa2}&\dots&\mathcal{L}_{\kappa\kappa}
 \end{array}
\right].
\end{equation}
Here, $\mathcal{L}_{\ell \ell}\in R^{\theta_\ell\times\theta_\ell}$ with $\sum_{\ell=1}^\kappa\theta_\ell=n$, and the directed subgraph associated with $\mathcal{L}_{11}$ is strongly connected, while $\mathcal{L}_{\ell\ell}$ with $\ell=2,\dots,\kappa$ are nonsingular $\mathcal{M}$-matrices. If $\mathcal{L}$ does not have the form given by (\ref{Perron}), it is always possible to reorder the labels of the agents such that the Laplacian matrix of the resulting graph is in the form of (\ref{Perron}). By Lemma \ref{Lem1}, there exists a vector $\omega_{\theta_1}=[\omega_{\theta_1,1},\dots,\omega_{\theta_1,\theta_1}]$ with $\omega_{\theta_1,i}>0$ for all $i=1,\dots,\theta_1$ such that $\omega_{\theta_1}\mathcal{L}_{11}=0_{\theta_1}^{T}$. Considering the Lyapunov function $V_1=\sum_{i=1}^{\theta_1}\omega_{\theta_1,i}\int_0^{\eta_i}\varrho_i(s)ds$ for the agents associated with $\mathcal{L}_{11}$, we obtain that $\lim_{t\rightarrow\infty}(x_i(t)-x_j(t))=0$ and $\lim_{t\rightarrow\infty}v_i(t)=v_r$ for all $i,j=1,\dots,\theta_1$. Since $\eta_i$ is continuous, there is a finite $T_{\theta_1}\geq T$ such that $|\eta_i(t)|\leq \frac{2m}{3k_i}$ for all $t\geq T_{\theta_1}$ and $i=1,\dots,\theta_1$, indicating that $\dot{x}_i=v_r-k_i\eta_i$ for all $t\geq T_{\theta_1}$. Define the relative error vector $\tilde{\chi}_1=[x_1-x_2,x_2-x_3,\dots,x_{\theta_1-1}-x_{\theta_1}]^{T}\in R^{\theta_1-1}$, whose dynamics satisfies $t\geq T_{\theta_1}$
\begin{equation}\label{d_x_tilde}
\dot{\tilde{\chi}}_1=-\Omega\tilde{\chi}_1,
\end{equation}
where $\Omega$ is a constant matrix that can be determined from $\mathcal{L}_{11}$. Following \cite[Thm. 2.14]{renbook2008distributed}, we obtain that system (\ref{d_x_tilde}) is  uniformly asymptotically stable. Therefore, there exist positive constants $q_1$ and $\varpi_1$ such that $(\varrho_i(\eta_i))^2=(k_i\eta_i(t))^2\leq q_1 e^{-\varpi_1t}$ for all $t\geq T_{\theta_1}$ and $i=1,\dots,\theta_1$.

We then consider the consensus for the agents associated with $\mathcal{L}_{22}$. Since $\mathcal{L}_{22}$ is a nonsingular $\mathcal{M}$-matrix, there exists a diagonal matrix $W_2=\diag\{\omega_{\theta_2,1},\dots,\omega_{\theta_2,\theta_2}\}$ with $\omega_{\theta_2,i}>0$ for all $i=1,\dots,\theta_2$ such that $G_2=W_2\mathcal{L}_{22}+\mathcal{L}_{22}^{T}W_2$ is positive definite \cite{mei2016distributed}. Considering $V_2=\sum_{i=\theta_1+1}^{\theta_1+\theta_2}\omega_{\theta_2,i-\theta_1}\int_0^{\eta_i}\varrho_i(s)ds$, we obtain that for all $t\geq T$
\begin{equation}\label{d_V_2}
\begin{array}{rcl}
\dot{V}_2&=&-\bar{\varrho}_2^{T}W_2\mathcal{L}_{21}\bar{\varrho}_1
-\bar{\varrho}_2^{T}W_2\mathcal{L}_{22}\bar{\varrho}_2\\
&=&-\bar{\varrho}_2^{T}W_2\mathcal{L}_{21}\bar{\varrho}_1-\frac{1}{2}\bar{\varrho}_2^{T}G_2\bar{\varrho}_2
\\
&\leq&-\frac{\lambda_{\min}(G_2)}{2}\|\bar{\varrho}_2\|^2+\|W_2\mathcal{L}_{21}\bar{\varrho}_1\|\|\bar{\varrho}_2\|
\\
&\leq&-\frac{\lambda_{\min}(G_2)}{4}\|\bar{\varrho}_2\|^2+\frac{1}{\lambda_{\min}(G_2)}\|W_2\mathcal{L}_{21}\bar{\varrho}_1\|^2,
\end{array}
\end{equation}
where $\bar{\varrho}_1=[\varrho_{1}(\eta_1),\dots,\varrho_{\theta_1}(\eta_{\theta_1})]^{T}$, $\bar{\varrho}_2=[\varrho_{\theta_1+1}(\eta_{\theta_1+1}),$ $\dots,\varrho_{\theta_1+\theta_2}(\eta_{\theta_1+\theta_2})]^{T}$, and $\lambda_{\min}(G_2)$ denotes the minimum eigenvalue of $G_2$. Recalling $(\varrho_i(\xi_i))^2\leq q_1e^{-\varpi_1 t}$ for all $t\geq T_{\theta_1}$ and $i=1,\dots,\theta_1$, we conclude that $\int_{T_{\theta_1}}^t\|W_2\mathcal{L}_{21}\bar{\varrho}_1(s)\|^2ds$ is bounded. Integrating (\ref{d_V_2}) over $[T_{\theta_1},t]$, we obtain that $\int_{T_{\theta_1}}^t\|\bar{\varrho}_2(s)\|^2ds$ is bounded. Note that $\|\bar{\varrho}_2(t)\|$ and $\|\dot{\bar{\varrho}}_2(t)\|$ are always bounded. It follows by direct application of Barbalat's lemma that $\lim_{t\rightarrow\infty}\varrho_i(\eta_i(t))=0$ for all $i=\theta_1+1,\dots,\theta_1+\theta_2$. Thus, we obtain that $\lim_{t\rightarrow\infty}\eta_i(t)=0$, $\lim_{t\rightarrow\infty}v_i(t)=v_r$, and $(\varrho_i(\eta_i))^2\leq q_2e^{-\varpi_2 t}$ for all $t\geq T_{\theta_2}$, $i=\theta_1+1,\dots,\theta_1+\theta_2$ with some positive constants $q_2$, $\varpi_2$, and $T_{\theta_2}$. Since $\mathcal{L}_{\ell\ell}$ are nonsingular $\mathcal{M}$-matrices, $\ell=3,\dots,\kappa$, applying the same steps as above, we conclude that $\lim_{t\rightarrow\infty}\eta_i(t)=0$ and $\lim_{t\rightarrow\infty}v_i(t)=v_r$ for all $i=\theta_2+1,\dots,n$, which together with \cite[Lemma 2.10]{renbook2008distributed} indicates that consensus among the multiagent system is achieved.

\begin{Rem}
The control of second-order multiagent systems with velocity and input constraints has been well researched in seminal work \cite{fu2019consensus}. However, the results may not be directly applicable in scenarios with agent uncertainties. To address this challenge, we propose a new framework that technically integrates saturated functions with fixed-time control and draws upon the techniques from \cite{fu2019consensus}, such as the construction of a Lyapunov function candidate. Our framework guarantees asymptotic consensus even in the presence of uncertainties in agent dynamics and the lack of neighboring velocity measurements. Furthermore, we extend the proposed framework to directed graphs with a spanning tree, providing a more comprehensive solution than previous methods with strongly connected graphs. 
As a result, this paper is a valuable supplement to the current literature.  
\end{Rem}

\section{Application to robotic manipulators}
In this section, we apply the distributed robust control framework in Section \ref{sec3} to address a consensus problem for single-link robotic manipulators with constrained velocities and torques. The manipulator can be described by the following dynamic model
\begin{equation}\label{manipulator}
\begin{array}{cll}
\dot{x}_{i}&=&v_i, \\
\dot{v}_{i}&=&I_i^{-1}(u_i-B_iv_i-M_i\g l_i\sin(x_i)),
\end{array}
\end{equation}
where $x_i\in R$, $v_i\in R$, and $u_i\in R$ represent the generalized position, generalized velocity, and input motor torque, respectively, $I_i$ is the total inertia of the link and the motor, $B_i$ is the damping coefficient, $M_i$ is the total mass, $\g$ is the gravitational acceleration, and $l_i$ is the distance from the joint axis to the link center of mass. By defining $b_i=1/I_i$, $\phi_i=B_i/I_i$, and $\tau_i=(M_i\g l_i)/I_i$, the dynamics in (\ref{manipulator}) can be equivalently written as: $\dot{v}_{i}=b_iu_i-\phi_iv_i-\tau_i\sin(x_i).$
We design a control torque $u_i$ that allows all manipulators to reach consensus on the generalized position, i.e., $x_i(t)-x_j(t) \rightarrow 0$ as $t\rightarrow\infty$ for all $i,j=1,\dots,n$, while guaranteeing that the velocity and input constraints given by (\ref{constraints_2}) are always satisfied. 
To achieve this objective, the following assumption is required.

\begin{Asu}\label{Asu3}
There exist some known positive constants $b_{\min}$, $\phi_{\max}$, and $\tau_{\max}$ such that $b_i\geq b_{\min}$, $\phi_i\leq\phi_{\max}$, and $\tau_i\leq \tau_{\max}$. Furthermore, $b_{\min}$ and $\tau_{\max}$ need to satisfy $b_{\min}u_{\max}-\tau_{\max}-\phi_{\max}\bar{v}>0$ to ensure the controllability of the agent, where $\bar{v}=\max\{|v_{\max}|,|v_{\min}|\}$.
\end{Asu}

The proposed distributed robust control law (\ref{u}) with (\ref{r_new}) can be applied to achieve the control objective by choosing the parameters $m$, $\alpha_i$, $z_i$, and $k_i$ to satisfy
\begin{eqnarray}\label{parameter_new2}
 &m<(v_{\max}-v_{\min})/2, z_i\leq (v_{\max}-v_{\min})/2-m,\nonumber\\
&(\tau_{\max}+\phi_{\max}\bar{v})/b_{\min}<\alpha_i<u_{\max},\nonumber\\
 &k_i<(b_{\min}\alpha_i-\tau_{\max}-\phi_{\max}\bar{v})/(d_i(v_{\max}-v_{\min})).
\end{eqnarray}
The result below can be established by mimicking the arguments presented in Theorems \ref{Thm1} and \ref{Thm2}, and thus its proof is omitted for simplicity.
\begin{Thm}\label{Thm3}
Consider a multiagent system of $n$ manipulators (\ref{manipulator}) satisfying Assumptions \ref{Asu1}-\ref{Asu3}. The proposed distributed robust control law (\ref{u}) with (\ref{r_new}) and (\ref{parameter_new2}) ensures that the manipulators achieve asymptotic consensus and the velocity converges to $v_r$, i.e., $\lim_{t\rightarrow\infty}(x_i(t)-x_j(t))=0$ and $\lim_{t\rightarrow\infty}v_i(t)=v_r$ for all $i,j=1,\dots, n$, and that the constraints of agent velocity and control torque given by (\ref{constraints_2}) are never violated.
\end{Thm} 
\section{Simulation study}
\begin{figure}
\begin{center}
\includegraphics[height=2.6cm]{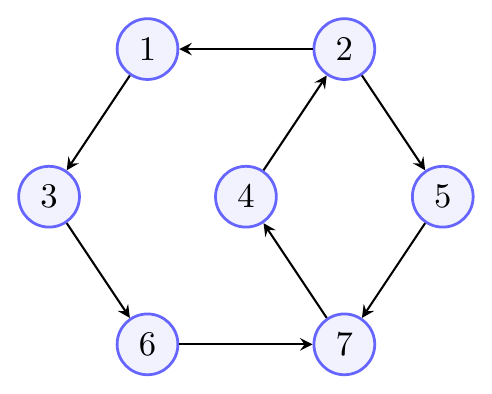}    
\caption{Directed interaction topology.} 
\label{fig_topology}                                 
\end{center}                                 
\end{figure}
\begin{figure}
\begin{center}
\includegraphics[height=6.5cm]{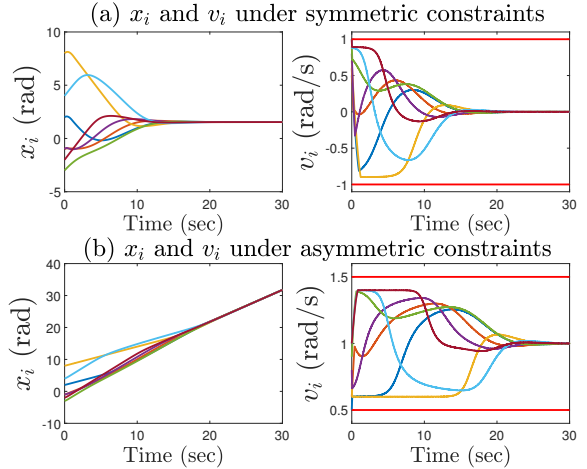}    
\caption{Trajectories of $x_i$ and $v_i$ under symmetric and asymmetric constraints.}  
\label{fig_x2}                                 
\end{center}                                 
\end{figure}

To validate the proposed theoretical results, we consider a set of 7 robotic manipulators described by (\ref{manipulator}), whose interaction topology satisfying Assumption \ref{Asu1} is given in Fig. \ref{fig_topology}. The manipulator parameters for our simulation scenario are $b_i=1$, $\phi_i=0.8$, and $\tau_i=0.5$ for $i=1,\dots,7$. The control torque is required to satisfy $|u_i(t)|\leq 2$. Regarding the velocity constraints, we consider the following two cases: (a) symmetric constraints: $-1\leq v_i(t)\leq 1$ and (b) asymmetric constraints: $0.5\leq v_i(t)\leq 1.5.$ According to (\ref{parameter_new2}), the control parameters are selected as $\gamma_i=1.5$, $\alpha_i=1.8$, $k_i=0.5$, and $z_i=0.1$. $m$ is set to $0.9$ for the symmetric constraint case and to $0.4$ for the asymmetric constraint case. The profiles of each manipulator position and velocity are exhibited in Fig. \ref{fig_x2}, from which we can observe that all the manipulators achieve consensus and the requirement on the velocity constraints is satisfied at all times.
\section{Conclusion}
The robust consensus control framework established in this work advances existing results on distributed consensus control in several directions. In addition to considering both the velocity and control input constraints, a generic class of agent dynamics has been investigated, accounting for system uncertainty and disturbance. Furthermore, our control law does not rely on the complete state information of neighboring agents and global information such as the eigenvalues of the (asymmetric) Laplacian matrix but only on local available position measurements such that asymptotic consensus in the multiagent system can be attained. 
Ongoing work implements the developed framework into quadrotor swarms.
\bibliographystyle{ieeetr}
\bibliography{autosam}

\begin{thebibliography}{10}

\bibitem{francis2016flocking}
B.~A. Francis and M.~Maggiore, {\em Flocking and rendezvous in distributed
  robotics}.
\newblock New York: Springer, 2016.

\bibitem{mei2016distributed}
J.~Mei, W.~Ren, and J.~Chen, ``Distributed consensus of second-order
  multi-agent systems with heterogeneous unknown inertias and control gains
  under a directed graph,'' {\em IEEE Transactions on Automatic Control},
  vol.~61, no.~8, pp.~2019--2034, 2016.

\bibitem{wang2019distributed}
G.~Wang, ``Distributed control of higher-order nonlinear multi-agent systems
  with unknown non-identical control directions under general directed
  graphs,'' {\em Automatica}, vol.~110, p.~108559, 2019.

\bibitem{ren2008consensus}
W.~Ren, ``On consensus algorithms for double-integrator dynamics,'' {\em IEEE
  Transactions on Automatic Control}, vol.~53, no.~6, pp.~1503--1509, 2008.

\bibitem{abdessameud2010consensus}
A.~Abdessameud and A.~Tayebi, ``On consensus algorithms for double-integrator
  dynamics without velocity measurements and with input constraints,'' {\em
  Systems \& Control Letters}, vol.~59, no.~12, pp.~812--821, 2010.

\bibitem{yi2019distributed}
X.~Yi, T.~Yang, J.~Wu, and K.~H. Johansson, ``Distributed event-triggered
  control for global consensus of multi-agent systems with input saturation,''
  {\em Automatica}, vol.~100, pp.~1--9, 2019.

\bibitem{saberi2022synchronization}
A.~Saberi, A.~A. Stoorvogel, M.~Zhang, and P.~Sannuti, {\em Synchronization of
  Multi-Agent Systems in the Presence of Disturbances and Delays}.
\newblock Cham: Birkh{\"a}user, 2022.

\bibitem{liu2018passivity}
Z.~Liu, A.~Saberi, A.~A. Stoorvogel, and M.~Zhang, ``Passivity-based state
  synchronization of homogeneous multiagent systems via static protocol in the
  presence of input saturation,'' {\em International Journal of Robust and
  Nonlinear Control}, vol.~28, no.~7, pp.~2720--2741, 2018.

\bibitem{yang2014global}
T.~Yang, Z.~Meng, D.~V. Dimarogonas, and K.~H. Johansson, ``Global consensus
  for discrete-time multi-agent systems with input saturation constraints,''
  {\em Automatica}, vol.~50, no.~2, pp.~499--506, 2014.

\bibitem{yang2016periodic}
T.~Yang, Z.~Meng, D.~V. Dimarogonas, and K.~H. Johansson, ``Periodic behaviors
  for discrete-time second-order multiagent systems with input saturation
  constraints,'' {\em IEEE Transactions on Circuits and Systems II: Express
  Briefs}, vol.~63, no.~7, pp.~663--667, 2016.

\bibitem{zhang2020semi}
M.~Zhang, A.~Saberi, and A.~A. Stoorvogel, ``Semi-global state synchronization
  for multi-agent systems subject to actuator saturation and unknown nonuniform
  input delay,'' {\em IEEE Transactions on Network Science and Engineering},
  vol.~8, no.~1, pp.~488--497, 2021.

\bibitem{lin2017distributed}
P.~Lin, W.~Ren, C.~Yang, and W.~Gui, ``Distributed consensus of second-order
  multiagent systems with nonconvex velocity and control input constraints,''
  {\em IEEE Transactions on Automatic Control}, vol.~63, no.~4, pp.~1171--1176,
  2018.

\bibitem{fu2019consensus}
J.~Fu, G.~Wen, W.~Yu, T.~Huang, and X.~Yu, ``Consensus of second-order
  multiagent systems with both velocity and input constraints,'' {\em IEEE
  Transactions on Industrial Electronics}, vol.~66, no.~10, pp.~7946--7955,
  2019.

\bibitem{zuo2020distributed}
Z.~Zuo, M.~Defoort, B.~Tian, and Z.~Ding, ``Distributed consensus observer for
  multiagent systems with high-order integrator dynamics,'' {\em IEEE
  Transactions on Automatic Control}, vol.~65, no.~4, pp.~1771--1778, 2020.

\bibitem{renbook2008distributed}
W.~Ren and R.~W. Beard, {\em Distributed consensus in multi-vehicle cooperative
  control}.
\newblock New York: Springer, 2008.

\end{thebibliography}
\end{document}